\documentstyle[aps,pre,epsfig,psfig,multicol]{revtex} 
\begin{document}                                                          

\titlepage
\title{
New approach to Monte Carlo calculation of buckling of
supercoiled DNA loops
}
\author{Yang Zhang}
\address{Institute of Theoretical Physics, 
Academia Sinica, P.O. Box 2735, Beijing 100080, China}
\maketitle    
    
\begin{abstract}

The short supercoiled circular
DNA molecules are shown to be glassy systems and
canonical Metropolis Monte Carlo simulations of the systems
tend to get stuck in local metastable 
energy basins.
A novel Monte Carlo algorithm is developed to alleviate the problem
of ``ergodicity breaking'' of the glassy systems, 
in which the Markov process is driven
by an explicitly analytic weight factor with enhanced probability in both
low- and high-energy regions. 
To characterize the degree of puckering of the supercoiled DNA loops,
a new quantity of aplanarity is introduced as the shortest principal axis
of configurational ellipsoid of DNA.
With the suggested Monte Carlo method,
the quantitative correlation between 
supercoiling degree and buckling of DNA is attained.
With supercoiling stress increasing, the conformational transition
from a circle to mono-, diplo- or triple interwound
superhelical structure will
take place in a successive but decreasingly abrupt mode.

\end{abstract}

\begin{multicols}{2}

The onset of supercoiling and supercoiling transitions of closed
circular deoxyribonucleic acid (DNA) are intriguing
phenomena from both a theoretical and a biological viewpoint.
It is generally believed that proper supercoiling strain is crucial
for the DNA molecule to pack and fold into chromosomes or to interact
with specific protein \cite{stryer1995}.
In the simplest mechanical description, in which one neglects all molecular 
inhomogeneities, a double-helical DNA molecule can be viewed as an
elastic rod, the elastic energy of which can be decomposed into
bending and torsional parts\cite{landou1986}. 
These two energy terms are usually decoupled since the bending energy
depends on local curvature of the center axis, while torsional energy
is the function of displacements of twist of the DNA basepairs
from the equilibrium positions. 
For a torsionally relaxed small loop, 
(e.g., nicked DNA polymer with one broken strand), 
the torsional energy of DNA vanishes, and DNA
molecule has the minimum bending energy of a flat circle.
The effective coupling between bending and torsional energies
rises for covalently closed DNA polymers, since the number of times two
strands of the DNA duplex are interwound, i.e., the  linking number
$Lk$, is of topological invariance.
The planar circle would no longer stand as a stable minimum of the 
total elastic energy when the link number deficit/excess
is beyond some threshold which is a function of bending and 
torsional stiffness of DNA polymer \cite{lebret1979}. 
In fact, the spatial configuration of circular DNA
depends upon the competition between bending and twist energies.
Planar DNA microcircles will pucker if the decrease in torsional energy
through changing the twist of the basepairs exceeds the increase of
bending energy caused by writhing of the polymer axis.

The buckling of short supercoiled DNA loops of $1000$ basepairs
(bp) was computationally studied by Schlick {\it et al.} with
deterministic techniques of energy
minimization and molecular dynamics (MD) \cite{schlick1992}. 
A catastrophic buckling transition from the circle to the conformation 
of figure-8 was presented with increasing supercoiling strain of DNA,
coinciding with earlier analytical calculations \cite{lebret1979}.
Metropolis Monte Carlo (MC) approach to
this buckling transition of DNA circle of 
$468$ bp was performed by Gebe and Schurr,
through tracing the writhing of the molecule \cite{gebe1996}.

\begin{figure}
\narrowtext
\centerline{\psfig{file=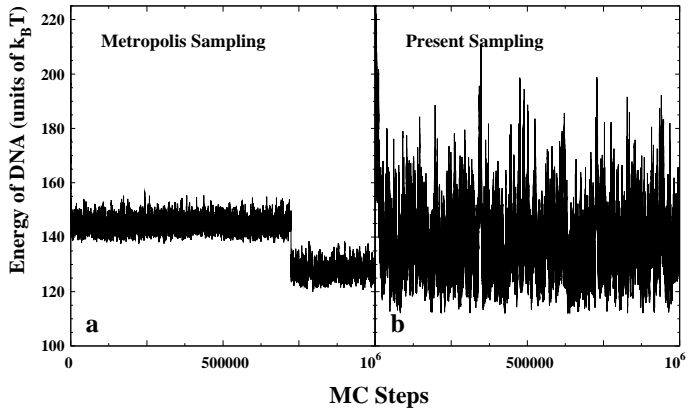,width=10cm}}
\caption{
The energy of DNA loop of 168 basepairs with linking
number deficit $\Delta Lk=5$, as a function of computer time:
({\bf a}), for canonical Metropolis method, the simulation is getting
stuck in a local energy basin at first 700,000 steps and
in another local energy minimum afterwards;
({\bf b}), for present sampling with new weight factor of
Eq. (\ref{weight}), much larger fluctuations are implemented,
which keeps the simulation from getting
trapped in local energy basins.
}
\label{fig1}
\end{figure}

In this Rapid Communication, we study the buckling dynamics of DNA
loops of $168$ bp through a new defined puckering degree (see below),
by Monte Carlo simulation of discrete wormlike chain model
\cite{vologodskii1992}.
The deformation energy of DNA rod is then approximated by 
the harmonic bending and twisting components.
In each update, an interval 
subchain containing arbitrary amount of links
is rotated around the straight line connecting the 
vertices bounding the subchain. 
The rotating angle is randomly taken in a interval chosen so that
about half of the proposed moves are accepted.
The excluded-volume effects and electrostatic repulsive interaction 
are incorporated by a hard-cylinder potential with an effective
diameter ($d>2$ nm) in the way the free energy of DNA will be
infinite when the distance of any non-adjacent parts of DNA is 
less than $d$.
However, as shown in Fig. \ref{fig1}a, the conventional Metropolis
simulation\cite{metropolis1953} of short DNA polymer of $168$ bp
with Boltzmann weight factor
tends to get stuck in some configurations with local metastable
minimum energies.

In fact, the energy landscape of supercoiled DNA chain is 
charactered by numerous local minima separated by energy 
barriers. 
At length scales comparable to the double-helix repeat of 
$3.4$ nm ($\approx 10.5$ basepairs) or the diameter of 
$2.0$ nm \cite{zhou1999}, the pairing and stacking enthalpy of the 
bases make the polymer remarkably rigid and the energy barriers 
significant high compared with Boltzmann weight factor. 
Since the probability of a canonical Metropolis procedure
to cross the energy barrier of height $\Delta E$ is proportional 
to $\exp(-\Delta E/k_BT)$, the simulations therefore
tend to get trapped in some local energy basins (Fig. \ref{fig1}a),
although Metropolis
Monte Carlo sampling is usually more effective than molecule
dynamics simulation for conformational changes which jumps to
different area of phase space \cite{leach1996}.
For the glassy systems such as short DNA loops,
different MC and MD simulations within finite CPU time and
sweeps may get stuck in different energy basins,
which renders the calculations of physical quantities
unreliable. 
One of the main aims of present work is at alleviating this problem of
``ergodicity breaking (EB)'' in Monte Carlo simulations of
glassy systems.

Technically, the above-mentioned EB problem is due to the fact 
that Boltzmann weight factor drops too quickly
(exponentially) with system energy $E$.
Therefore, the probability of the Markov process
to jump over the high energy barriers is exponentially damped. 
In the following, we consider a new weight factor
\begin{equation}
\label{weight}
w(E)=e^{-E+(\sqrt{2}/\sigma)|E-\langle E\rangle |},
\end{equation}
where $\langle E\rangle$ is the averaged energy of system,
and $\sigma^2$ ($=n_F/2$) is the mean squared deviation of energy
of the canonical thermodynamic system, and $n_F$
the number of degree of freedom.
For the discrete closed wormlike-chain with $N$ links,
the number of degree of freedom is $n_F=2N-6$.
Here and after the energy and rigidity parameters are
scaled by $k_BT$, so the Boltzman inverse
temperature parameter $\beta=1/k_BT$
is omitted in our equations.

With the introduction of factor of 
$\exp[(\sqrt{2}/\sigma) |E-\langle E\rangle |]$,
the sharp peak of energy distribution of canonical ensemble 
is damped and the important sampling in low-energy region 
is reinforced. 
Especially, within the new weight factor of Eq. (\ref{weight}), the
probability in higher-energy region
is enhanced exponentially, which makes the simulation escape from
local energy basins easily.
In Fig. \ref{fig1}b we present the MC time series of energy of the same
DNA system 
but with the new weight factor of Eq. (\ref{weight}).
Indeed, the simulation of the new artificial ensemble
covers a much wider energy range than that of the canonical run, 
which efficiently keeps the simulation from getting stuck in local
minima.

\begin{figure}
\centerline{\psfig{file=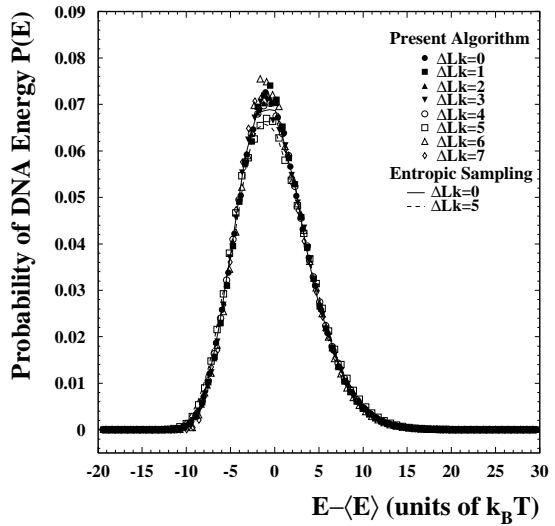,width=8cm}}
\caption{
The probability distributions of energies of DNA loops,
which is calculated by reweighting 
the artificial ensemble with Eq. (\ref{reweight}).
After translating the energy to the averaged value $\langle E\rangle$,
the distributions of different linking number deficit $\Delta Lk$ 
fall on the same curve.
The results of the simulation by entropic sampling 
method{\protect\cite{berg1992,hansmann1993,newman1999,lee1993}}
coincide with that by present approach, 
but its CPU time is twice more than the latter.
}
\label{fig2}
\end{figure}

In the artificial ensemble of (\ref{weight}), 
each configuration of energy $E$ of DNA loop,
in fact, represents 
$n(E)$ ($=\exp[-(\sqrt{2}/\sigma) |E-\langle E\rangle |]$)
ones of the real thermodynamic system.
Through reweighting the artificial sample \cite{fosdick1963}, i.e.
\begin{equation}
\label{reweight}
\langle A\rangle ={\sum_{i=1}^{N_{\rm sweep}}A(E_i)
e^{-{\sqrt{2}\over\sigma} |E_i-\langle E\rangle |}
\over\sum_{i=1}^{N_{\rm sweep}}
e^{-{\sqrt{2}\over\sigma} |E_i-\langle E\rangle |}},
\end{equation}
where $N_{\rm sweep}$ is the number of MC sweeps of the artificial
sample, we can 
obtain the expectation value or probability distribution
of the considered quantity $A$ in the
real physical system.
As an illustrated example, 
we show in Fig. 2 the probability distributions of 
DNA energies, which are calculated from the
artificial sample of 2,000,000 MC runs.
As expected, after the translation of the DNA energies to their averaged
values $\langle E\rangle$, all the distributions with different
supercoiling strains precisely fall onto the same curve.
In fact, this scaling behavior of the energy distribution of
canonical ensemble is the foundation that
we could use the same compensative factor in Eq. (\ref{weight}) to
alleviate the EB problem in different energy
systems. 

It should be mentioned that the EB difficulty
has been longly standing in the Monte Carlo simulations of
glassy systems.
The most often-used technique of earlier dealing with this problem
in first-order phase transitions \cite{berg1992}, 
protein folding problem \cite{hansmann1993}
and other glassy systems \cite{newman1999},
is entropic sampling method\cite{lee1993,note},
in which
the spectral density of the considered system is calculated numerically 
by an iterative procedure so that the simulation can be performed as a
random walk in a desired range of phase space. 
Depending on the rate of convergence of iteration and the size of
simulated energy range,
the determination of weight factor is usually 
tedious and very much
time-consuming\cite{berg1992,hansmann1993,newman1999,lee1993}.
To get an approximate flat distribution of energy spectral 
in the simulation of 
our short DNA loops, for example,
we need to spend more than 50\% of the whole CUP time
to achieve this task.
In Fig. 2 we also show the probability distribution of DNA energies
of the canonical ensemble obtained from the entropic sampling technique
for $\Delta Lk=0$ and $5$ respectively.  The results coincide
with that by the present method.

The only parameter to be determined in Eq. (\ref{weight})
is the averaged energy of the system.
In principle, such estimates can be found in an simple iterative 
way \cite{hansmann1997}.
One first sets an initial guess of the averaged energy 
and performs a simulation with small number of Monte Carlo sweeps, 
and gets a new value of the averaged energy 
which is a better estimator for the real energy.
One can run the simulation using the new estimator of $\langle E\rangle$
to get a newer one and iterate this process until $\langle
E\rangle\approx\sqrt{\langle E^2\rangle -n_F/2}$. 
According to our calculations, the averaged energy converges to the
stable value very quickly, and just a few times of iterations are enough to
get a precise estimator of the averaged energy. 
In Table I is listed the averaged energies of the DNA loops of some
different supercoiling restrains, which are obtained by 3 iterations.
The estimation of averaged energy can be further facilitated by
information of its ground state value $E_0$. For example, the
ground state of DNA loop for $\Delta Lk=0$ corresponds to the flat
circle, energy of which is $18.14k_BT$ in our system. 
So we can directly calculate the
averaged energy by $\langle E\rangle =E_0+n_F/2
=35.14 k_BT$.
\end{multicols}

\widetext
\begin{table}
\label{table1}
\caption{
The averaged energies of 168-basepair-DNA loops
for different supercoiling restrains,
which are obtained by 3 iterations
as demonstrated in the text.
}
\begin{center}
\begin{tabular}{cccccccccccccccc}
$\Delta Lk$ &
0.0 &
0.2 &
0.4 &
0.6 &
0.8 &
1.0 &
1.2 &
1.4 &
1.6 &
1.8 &
2.0 &
2.5 &
3.0 &
3.5 &
4.0 
 \\
 $\langle E\rangle\  ( {\rm in}\ k_BT)$&
35.09&
36.03 &
39.11 &
44.15 &
51.11 &
60.00 &
69.78 &
73.80 &
78.84 &
85.10 &
91.58&
107.84&
127.47&
150.38&
177.78
\end{tabular}
\end{center}
\end{table}

\begin{multicols}{2}
In previous literatures, e.g. Refs. \cite{gebe1996,vologodskii1992}, 
writhing
number ($Wr$) was often used to characterize
the tertiary structure and handedness of circular DNA,
which is defined as the difference between the linking number $Lk$
of two strands and the twisting number $Tw$ of basepairs
\cite{white1969}.
However, $Wr$ has some features which 
obviously hinder itself as the most proper definition of the puckering
degree of DNA loops.
For example, for a circular DNA wrapping around a sphere, it may be
considerably puckered, however, $Wr$ definitely equals to zero; 
when DNA takes a small displacement of passage through itself, 
the puckering degree and aplanarity should keep almost unchanged, 
however, $Wr$ discontinuously jumps. 

To give a proper definition of the degree of puckering of DNA loops, 
let us at first define a $3\times 3$ symmetric coordinate tensor:
\begin{equation}
\label{tensor}
T_{ab}={3\int(r_a(s)-r_{0a})(r_b(s)-r_{0b})ds\over\int({\bf r}(s)-
{\bf r}_0)^2ds}, \  a,b=1,2,3,
\end{equation}
where ${\bf r}(s)$ 
denotes the axis vector of DNA polymer along
with its arc-length $s$, ${\bf r}_0=(1/L)\int {\bf r}(s)ds$ the
center of mass of the
DNA polymer with total arc-length of $L$, 
$r_{a(b)}(s)$ the projection of the axis vector on $a(b)$-axis
of 3-D Cartesian coordinate system.
The positive definite tensor $T_{ab}$ has three eigenvalues $T_i$'s 
with $0\leq T_1\leq T_2\leq T_3$ and $T_1+T_2+T_3=3$. 
In fact, $T_{ab}$ represents the configurational ellipsoid of 
DNA polymer and $T_i$ its $i'$th major axis.  
So the smallest eigenvalue $T_1$ signifies the degree of
puckering of the DNA loop from a planar circle and we call it
``aplanarity of DNA loop''. 

In Fig. 3 is shown the major axes of DNA loop of $168$ bp with
different supercoiling stresses, which are calculated through the
reweighting equation (\ref{reweight}) after $2,000,000$ MC runs.
To access the handedness of the buckling,
we also present the
values of $Wr$ and $\Delta Tw$ versus $\Delta Lk$ in Fig. 3d.
With supercoiling strain increasing, the planar circle will become
unstable and a conformational transition from circle to figure-$8$
takes place at $\Delta Lk_c\sim -1.2$, which manifests itself 
as an abrupt jump in all data of major axes, 
writhing and twisting number, as well as bending and
twisting energies (data not shown).
This critical value of $\Delta Lk_c$ is in good
agreement with the analytical results \cite{lebret1979}, i.e.
$\Delta Lk_c=\sqrt 3A/C$, where bending persistent length $A$ and
twisting persistent length $C$ are taken as $53$ nm and $72.5$ nm
respectively in our simulations according to corresponding
experimental data \cite{horowitz1984}.
When $\Delta Lk$ continues to increase, the figure-8 configuration will 
become unstable again. DNA loop will take diplo- or triple interwound
superhelical conformation after $\Delta Lk$ is beyond about $-2.2$ or
$-3.4$,
which can be most clearly identified in the data of aplanarity
(see Fig. 3a).
However, the latter transitions are less abrupt than that
of circle to figure-8,
as denoted by the width of peaks in Fig. 3a.

\begin{figure}
\narrowtext
\centerline{\psfig{file=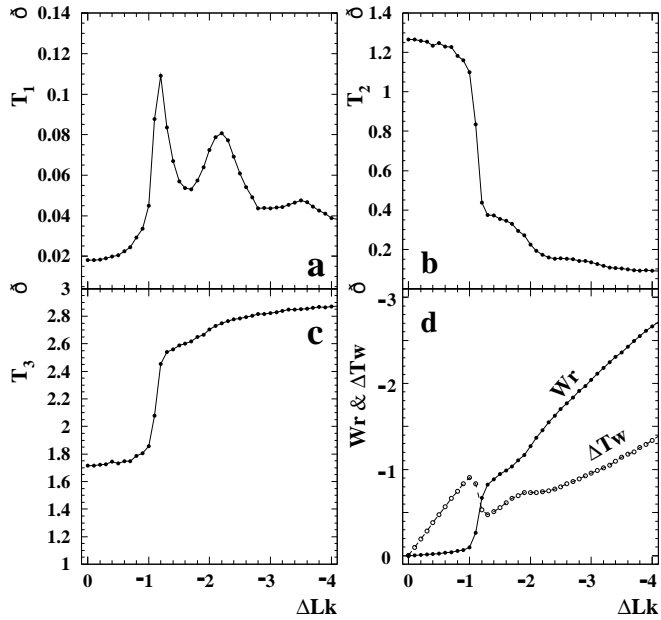,width=10cm}}
\caption{
Averaged major axes as well as writhing and twisting number 
as function of
linking difference of 168-basepair-DNA loop,
with statistic error bars of MC calculations inside the symbols.
The lines serve to guide the eye.
}
\label{fig3}
\end{figure}

In summary, 
we showed that short supercoiled DNA polymer
is a glassy system and canonical Metropolis simulation
tends to get
stuck in some local metastable energy basins.
A new Monte Carlo algorithm
with an explicitly analytic weight factor
is introduced to solve the problem of
ergodicity breaking,
in which
the thermalization is reinforced in both low- and high-energy regions.
Compared with earlier approach of entropic sampling,
the probability weight factor of present algorithm 
is clearly easier to be determined
and the implementation is about 2-fold CPU time-saving.
As a general Monte Carlo method,
the present approach can be also used in other
thermodynamic systems with frustration.

To characterize quantitatively the degree of puckering of the
circular DNA,
a new quantity of aplanarity has been introduced as 
the shortest major axis of configurational tensor of DNA.
With the developed Monte Carlo method,
a quantitative correlation between the buckling of DNA
loop of $168$ basepairs and supercoiling degree is presented.
Abrupt configurational transition from circle to figure-8 takes
place at the critical supercoiling stress of
$\Delta Lk_c$ ($\sim 1.2$), which is in good agreement
with previous analytical results.
With further increasing linking difference, DNA loops will take 
successively diplo- and triple superhelical conformation through the
decreasingly abrupt transitions. 

I am grateful to B. L. Hao, Z. C. Ou-yang and H. J. Zhou
for helpful discussions, and
U. H. E. Hansmann, S. Lindsay and J. M. Schurr for correspondence.
I also thank the anonymous referee for bringing the works of
Dr. Schurr's group to my attention. 
The project is partly sponsored by SRF for ROCS, SEM.

\end{multicols}

\end{document}